\documentclass[pra,twocolumn]{revtex4}

\usepackage{newlfont}
\usepackage{amssymb}
\usepackage{amsfonts}
\usepackage{amsmath}
\usepackage{graphicx}
\usepackage{bbm}

\newcommand{\be}{\begin{equation}}
\newcommand{\ee}{\end{equation}}
\newcommand{\bqa}{\begin{eqnarray}}
\newcommand{\eqa}{\end{eqnarray}}

\newcommand{\ket}[1]{|#1\rangle}
\newcommand{\bra}[1]{\langle#1|}
\newcommand{\prj}[1]{|#1\rangle\langle#1|}

\newcommand{\tr}{\textrm{Tr}}

\newcommand{\um}{\mbox{\bf 1}}

\begin{document}

\title{Concurrence via entanglement witnesses}

\author{Florian Mintert}
\affiliation{Department of Physics,
Harvard University, Cambridge, Massachusetts 02138, USA}

\begin{abstract}
We derive an experimentally observable lower bound on concurrence of mixed quantum states in terms of an entanglement witness,
relating measurements on single states with those on two copies.
\end{abstract}

\maketitle

Although entanglement does have unambiguously observable consequences,
such as the provision of quantum teleportation, entanglement {\it per se} is not an observable
in the strict quantum mechanical sense:
there is no observable, {\it i.e.} hermitean operator $E$,
such that its expectation value could define a valid entanglement monotone or measure.

Even though there is no unique measure of entanglement, and there is not even a generally accepted list of axioms that a measure of entanglement has to satisfy,
there is {\em the} fundamental property that
entanglement is invariant under local unitary operations.
That is, for a given state $\ket{\Psi_{0}}\in{\cal H}={\cal H}_{1}\otimes{\cal H}_{2}$,
all states $\ket{\Psi}={\cal U}_{1}\otimes{\cal U}_{2}\ket{\Psi_{0}}$,
with arbitrary unitary transformations ${\cal U}_{i}$ acting on ${\cal H}_{i}$, ($i=1,2$)
have exactly the same entanglement properties --
independent of the choice of measure.
Since this holds for arbitrary states $\ket{\Psi_{0}}$, and similarly also for mixed ones,
any observable $E$ that could potentially define a measure
also needs to have the same symmetry, {\it i.e.}
$E={\cal U}_{1}^{\dagger}\otimes{\cal U}_{2}^{\dagger}\ E\ {\cal U}_{1}\otimes{\cal U}_{2}$
for arbitrary local unitaries.
However, the only operator that enjoys this property is the identity $\um$,
{\it i.e.} the `trivial' observable that returns the same expectation value for {\em any} state.

Thus, in order to characterize entanglement in a laboratory experiment one needs to go beyond
measuring a single observable.
A commonly pursued approach is quantum state tomography \cite{RevModPhys.29.74}:
after measuring a complete set of observables one can reconstruct the actual quantum state,
and is left with the mathematical problem of evaluating some entanglement measure,
or check a separability criterion.
Although this has been implemented frequently and very successfully, {\it e.g.} in \cite{W8},
it is not completely satisfactory an approach.
Firstly, the number of observables to be measured grows rapidly with the system size,
so that for practical reasons tomography is viable only for small systems;
and secondly, one would expect that entanglement -- providing one of the most striking differences between classics and quantum -- has some more direct consequences to be observed.
And indeed, entanglement witnesses \cite{hor97} define an observable that gives a much more direct experimental insight to the separability properties of a given state \cite{witness04}.
However, whereas witnesses allow to distinguish separable from entangled states,
they do not provide a quantitative description, that is, they do not define an entanglement measure.
Though, recently some techniques have been developed to use witnesses in order to find bounds on entanglement measures
\cite{brandao:022310,quant-ph/0605155,quantph0607163,quantph0607167}.

A different approach is to not measure only on a single quantum state.
Since a state has to be prepared repeatedly for reliable measurement statistics anyway,
one can also wait until a state has been prepared twice, or prepare it twice at the same time,
and then measure collective properties of this twofold copy
\cite{ekert:217901,toddbrun,exp_mc,quantph0604109,quant-ph/0606017}.
In such a fashion one finds observables that are invariant under local unitaries,
and, indeed, the concurrence of pure states is then given as a regular observable,
and similarly, simple observable bounds are available for mixed states \cite{mc_mixed}.
In the present contribution we would like to point out some relations between the apparently different approaches of measuring either on a single copy or on two of them.

The concurrence $c(\Psi)$ of a pure state $\ket{\Psi}$ can be defined via the expectation value
of the hermitean operator $P_{-}^{(1)}\otimes P_{-}^{(2)}$, where $P_{-}^{(i)}$ is the projector onto the antisymmetric subspace
of ${\cal H}_{i}\otimes{\cal H}_{i}$ \cite{physrep}.
And, for a mixed state, concurrence can be constructed as convex roof
$c(\varrho)=\inf\sum_{i}c(\psi_{i})$,
where the infimum is to be taken among all ensembles $\{\ket{\psi_{i}}\}$ of suitably normalized states that are in accordance with the density matrix $\varrho=\sum_{i}\prj{\psi_{i}}$.
Such an optimization procedure is an involved mathematical task,
and a general algebraic solution is known only for the smallest possible systems \cite{wot98};
for higher dimensional systems there are only lower bounds available \cite{lb,chen:040504}.
Moreover, neither the exact solution nor the bounds are directly measurable -- they can only be evaluated if the density matrix is known,
typically via quantum state tomography.
Yet, there exists also a measurable lower bound on concurrence, provided a measurement can be performed on a twofold copy of a state.
In terms of the above projectors onto antisymmetric spaces, and the symmetric counterparts
$P_{+}^{(i)}$, this bound reads \cite{mc_mixed}
\be
c(\varrho)^{2}\ge 4\ \tr(\varrho\otimes\varrho\ V)\ ,
\label{bound_copy}
\ee
with either of the two choices
$V=P_{-}^{(1)}\otimes(P_{-}^{(2)}-P_{+}^{(2)})$, and,
$V=(P_{-}^{(1)}-P_{+}^{(1)})\otimes P_{-}^{(2)}$.

The advantages and disadvantages of this bound as compared to those derived from measurements
on single copies are apparent:
Eq.~(\ref{bound_copy}) defines a single measurement prescription that is applicable to all states,
which is a consequence of the invariance of $P_{\mp}$ under local unitaries.
However, this ease comes at the expense of the required simultaneous availability of two copies of a quantum state.
If one has some a priori information on $\varrho$, and, if one can easily adjust the measurement apparatus to suit the state to be characterized,
then one might be willing to give up this invariance, and return to measurements on a single copy only.
For those cases, we discuss how the approach of collective measurements can be used to find
observables on a single copy of $\varrho$
that also provide a lower bound on concurrence.

In order to obtain Eq.~(\ref{bound_copy}) above, it was shown in \cite{mc_mixed} that the inequality
\be
c(\psi)c(\phi)\ge 4\bra{\psi}\otimes\bra{\phi}V\ket{\psi}\otimes\ket{\phi}
\label{ineq}
\ee
holds for two arbitrary pure states $\ket{\psi}$ and $\ket{\phi}$.
Now, consider two mixed states $\varrho$ and $\sigma$,
with some decomposition into pure states
$\varrho=\sum\ket{\psi_{i}}\bra{\psi_{i}}$ and
$\sigma=\sum\ket{\phi_{i}}\bra{\phi_{i}}$.
By virtue of Eq.~(\ref{ineq}), one obtains
\begin{eqnarray}
\bigl(\sum_ic(\psi_{i})\bigr)\bigl(\sum_{j}c(\phi_{j})\bigr)
&\ge&
4\sum_{ij} \bra{\psi_{i}}\otimes\bra{\phi_{j}}V\ket{\psi_{i}}\otimes\ket{\phi_{j}}\nonumber\\
&=&
4\ \tr(\varrho\otimes\sigma\ V)\ .
\end{eqnarray}
Since this holds for any decomposition of $\varrho$ and $\sigma$,
that is, in particular, for optimal ones
that achieve the infimum in the convex roof construction of concurrence,
this directly leads to
\be
c(\varrho)c(\sigma)\ge 4\ \tr(\varrho\otimes\sigma\ V)\ .
\label{diffstate}
\ee

Now, take $\varrho$ to be the state of interest,
the concurrence of which should be measured;
$\sigma$ does not need to be a state of a real quantum system
so that a measurement is performed on two systems.
It can also be considered a mathematical auxiliary quantity
with the help of which one can define the observable $W_{\sigma}=-4\tr_{2}(\um\otimes\sigma\ V)/c(\sigma)$,
where the partial trace is taken over the second copy of ${\cal H}$.
$W_{\sigma}$ is just a regular hermitean operator to be measured on a {\em single} copy of ${\cal H}$,
and it can also be defined -- without invoking a second copy of ${\cal H}$ --
in terms of $\sigma$ and its reduced density matrix $\sigma_{1}=\tr_{2}\hspace{.5mm}\sigma$:
$\tr_{2}(\um\otimes\sigma\ V)=(\sigma-\um\otimes\sigma_{2})/2$ for
$V=P_{-}^{(1)}\otimes(P_{-}^{(2)}-P_{+}^{(2)})$, and
similarly for $V=(P_{-}^{(1)}-P_{+}^{(1)})\otimes P_{-}^{(2)}$ \cite{PhysRevA.59.4206,PhysRevA.60.893}.

Now, as a direct consequence of Eq.~(\ref{diffstate}), one has the lower bound on concurrence
\be
c(\varrho)\ge-\tr(\varrho W_{\sigma})\ ,
\label{newbound}
\ee
in terms of the observable $W_{\sigma}$ that is an entanglement witness for any positive operator $\sigma$:
for any separable state $\varrho$,
the expectation value of $B_{\sigma}$ is non-negative, since $c(\varrho)$ vanishes.
Therefore, a negative expectation value of $W_{\sigma}$ unambigously characterizes a state $\varrho$ to be entangled.
However, via Eq.~(\ref{newbound}) such witnesses do not only qualitatively distinguish separable from entangled states,
but also provide a quantitative description.

Of course, for an arbitrary state $\sigma$ it can be very involved to determine $c(\sigma)$,
which is necessary to obtain $W_{\sigma}$.
But often pure states provide good witnesses,
and the concurrence of pure states is a simple algebraic function.
Therefore, the witness $W_{\sigma}$ can be found purely algebraically,
without any optimization to be performed.
This, in particular, facilitates an optimization of Eq.~(\ref{newbound}) over states $\sigma$,
such as to find a possibly large bound.
And even if $\sigma$ should be a mixed state, and if its concurrence is not known,
any upper bound on $c(\sigma)$ also provides a valid witness for Eq.~(\ref{newbound});
and {\em reliable} upper bounds can always be obtained, simply due to the convex roof construction,
where by definition the solution is an infimum.

Thus, the present approach leads to a constructive, algebraical definition of entanglement witnesses
that provide bounds on concurrence of mixed states,
and sheds some light on the interconnection between
individual properties of a single state, and collective properties of copies thereof.

Stimulating discussions with Marek Ku\'s, Andreas Buchleitner, Adam Wasserman, Adriano Arag\~ao and Leandro Aolita,
as well as
financial support by Alexander von Humboldt fundation are gratefully acknowledged.

\end{document}